\documentclass[lettersize,journal]{IEEEtran}
\usepackage[caption=false,font=normalsize,labelfont=sf,textfont=sf]{subfig}
\usepackage{textcomp}
\usepackage{stfloats}
\usepackage{url}
\usepackage{verbatim}
\usepackage{graphicx}
\graphicspath{{./}{./img/}{./img/plots_0529-2153_0707-1121/}} 
\DeclareGraphicsExtensions{.pdf,.jpeg,.png}
\usepackage{cite}
\usepackage{soul, xcolor}
\usepackage{booktabs}
\usepackage{multirow} 
\usepackage{adjustbox}
\begin{document}

\title{BenchFaaS: Benchmarking Serverless Functions in an Edge Computing Network Testbed}

\author{\IEEEauthorblockN{Francisco~Carpio,~Marc~Michalke~and~Admela~Jukan}
    \IEEEauthorblockA{\\ \textit{Institute of Computer and Network Engineering,}
        \textit{Technische Universit{\"a}t Braunschweig}, Germany \\
        E-mail: \{f.carpio, m.michalke, a.jukan\}@tu-bs.de}

}



\newcommand{\testbedname}{BenchFaaS} 

\newcommand{\smallvm}{\texttt{vm.small}}
\newcommand{\mediumvm}{\texttt{vm.medium}}
\newcommand{\largevm}{\texttt{vm.large}}
\newcommand{\rpmetal}{\texttt{rp.metal}}
\newcommand{\lcl}{\texttt{loc}}
\newcommand{\cld}{\texttt{cld}}
\newcommand{\wst}{\texttt{ewst}}
\newcommand{\typ}{\texttt{etyp}}
\newcommand{\opt}{\texttt{eopt}}

\maketitle

\begin{abstract}
    The serverless computing model has evolved as one of the key solutions in
    the cloud for fast autoscaling and capacity planning. In edge computing
    environments, however, the serverless model is challenged by the system
    heterogeneity and performance variability. In this paper, we introduce
    \testbedname, an open-source edge computing network testbed which automates
    the deployment and benchmarking of serverless functions. Our edge computing
    network considers a cluster of virtual machines and Raspberry Pis, and is
    designed to benchmark serverless functions under different hardware and
    network conditions. We measure and evaluate: (i) overhead incurred by
    testbed, (ii) performance of compute intensive tasks, (iii) impact of
    application payload size, (iv) scalability, and (v) performance of chained
    serverless functions. We share the lessons learnt in engineering and
    implementing the testbed. We present the measurement results and
    analyze the impact of networked infrastructure on serverless performance.
    The measurements indicate that a properly dimensioned edge computing network
    can effectively serve as a serverless infrastructure.

\end{abstract}


\section{Introduction}
The Function-as-a-Service (FaaS) computing model removes decision-making on
scaling thresholds, reduces costs by charging only when applications are
triggered, and reduces application starting times \cite{Castro2019}. Serverless,
the execution model that FaaS implements, has been well studied in the cloud
context \cite{martins2020}. Serverless furthermore adopts the most recent
advances in containerization technologies, not only in the cloud, but also in
edge computing \cite{Morabito2017, Palade2019}. The serverless concept is
especially attractive to application developers in a combined edge and cloud
scenario to ensuring low latency together with high computing efficiency
\cite{aske2018}. Today, even the computationally intensive applications,
including deep learning models, run in edge computing systems \cite{Wang2020}.

From the networking perspective, it is critical to recognize that
not only hardware appliances can impose limitations on serverless functions, but
their placement in the network also plays a significant role in system
performance. This is a vastly unexplored area today, and comes in
addition to obvious and important known challenges, such as resource volatility
and constraints, as well as heterogeneity \cite{Caprolu2019}. In addition,
containerization and software abstractions of heterogeneous edge devices have
become essential to implement \cite{Morabito2018, Das2019a}. Especially the
so-called chaining of serverless functions take advantage of the modularity and
flexibility of containerization \cite{baldini2017}. To benchmark the performance
of chaining functions is particularly challenging, especially in the context of
the underlying network.

To address these challenges, we engineered \emph{\testbedname}, a programmable
and open-source edge computing network system testbed where users can run
serverless functions over a Kubernetes cluster of virtual machines (VMs) and
Raspberry Pis (RPs). The testbed can not only be configured to run on
heterogenous edge computing resources, placed in various locations in the
network, but the underlying network system can also be configured with various
properties typical to wide area network (WAN), including varying network delays,
variances and packet losses. We measure and evaluate: (i) overhead (or
performance offset) of the testbed, (ii) performance of compute intensive tasks,
(iii) impact of application payload size, (iv) scalability, and, (v) function
chaining.  The measurements indicate that a properly dimensioned system can
effectively deploy resource constrained edge computing devices as a serverless
network infrastructure.We show that resource dimensioning, and not
only the delay performance, becomes a key factor to consider when designing
serverless systems over edge computing networks. 

\emph{\testbedname} is offered as free and open source edge network
testbed\footnote{F. Carpio and M. Michalke "BenchFaaS" GitHub, 2022. [Online].
Available: https://github.com/fcarp10/benchfaas.} suitable for serverless
function performance benchmarking, and is fully reproducible. As such, it
complements the related community-driven experimental studies, focused on the
performance of different serverless platforms over third-party cloud/edge
services \cite{martins2020, Das2019a}. It also complements other related efforts
focused on open source serverless platforms however without consideration of the
underlying networked infrastructure \cite{junfeng,Palade2019}. Earlier versions
of \emph{\testbedname} were conceptualized in our previous work
\cite{carpio2021}.

The rest of the paper is organized as follows. Section II presents
the architectural principles behind the testbed design. Section III describes
the testbed configurations and benchmarking tests. Section IV analyzes the
measurements. Section V concludes the paper.

\section{\testbedname~Architectural Design}\label{sec:arch}

Building an edge computing system requires us to consider a heterogenous
infrastructure distributed over different network segments. On the one hand, we
need to support devices distributed over traditional LANs hidden behind NATs,
and appliances allocated in cloud network domains hidden behind strict
firewalls. On the other hand, the system needs to support heterogeneous devices;
we consider the two most used CPU architectures today, i.e., AMD64 and ARM64,
with the host operating system (OS) Linux-based and at least 2GB of RAM. We
propose a system architecture as shown in Fig. \ref{fig:arch}.  The remainder of
this section provides more implementation details of the architecture
deployment.

\subsection{Testbed Controller}

\begin{figure}[!t]
    \centering
    \includegraphics[width=1.0\columnwidth]{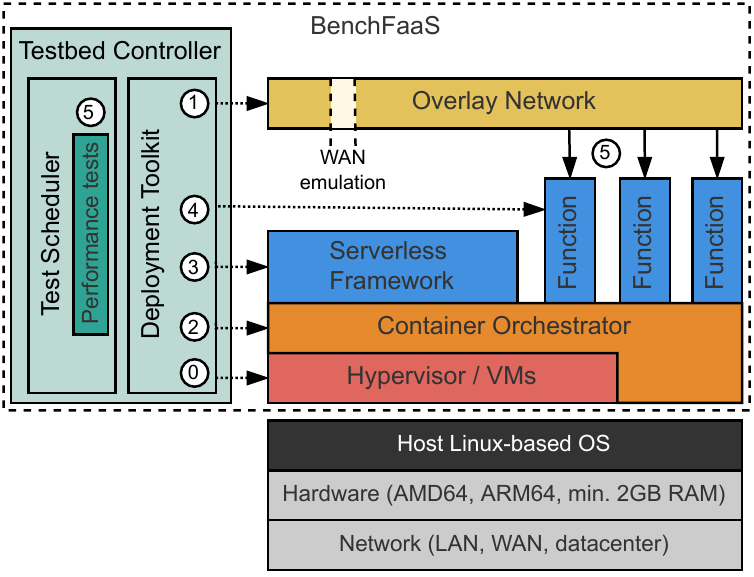}
    \caption{\testbedname~testbed architecture deployment}
    \label{fig:arch}
\end{figure}

The key component for \testbedname~users is \emph{Deployment Toolkit} (see Fig.
\ref{fig:arch}) which automates the deployment of the different components and
applies the specific programmable configuration of the testbed in five steps. In
\emph{step 0}, the testbed can be  configured to use VMs, otherwise the
functions can run over physical machines (PMs). In \emph{step 1}, the emulation
of the overlay network is configured, including emulation of delay, variance and
packet loss parameters. In \emph{step 2} and \emph{step 3} the deployment of the
container orchestrator and the installation of the serverless platform is
applied, respectively. Once this is done, in \emph{step 4}, all serverless
functions are deployed. Then, \emph{Test Scheduler}, in \emph{step 5}, performs
a series of benchmarking tests for serverless functions given a specific
infrastructure and network setup. Both \emph{Test Scheduler} and
\emph{Deployment Toolkit} are logically placed inside \emph{Testbed Controller},
which can reconfigure the parameters and run the aforementioned steps for
different network/hardware configuration setups. 

In this way, the \emph{Testbed Controller} is in charge of the deployment of all
the software components and of carrying out the performance tests for a specific
network and hardware setup; these tests are specific to the functions deployed
and are user-defined. The salient feature of this controller is that it
automates the deployment of all components over different network and hardware
configurations in order to analyze the serverless performance of the system over
different scenarios. All the code with specific instructions on how to configure
the testbed, including how to deploy it over VMs or PMs, is freely available at
the testbed github repository. 

\subsection{Overlay Network}

When connecting devices from different networks located behind NATs (Network
Address Translation) or firewalls, it is required to either apply manual NAT
configurations, firewall rules, or to use traditional privacy-oriented VPN
solutions (such as OpenVPN or Wireguard). While these VPN solutions can create
mesh VPN networks by tunneling all traffic, their configuration is laborious and
does not scale with an increased number of connected devices. Instead, tools
such as Nebula or Zerotier make use of easily scalable beaconing servers and UDP
hole punching techniques to directly interconnect devices over the network,
avoiding manual firewall configurations and creating, in this way, mesh overlay
networks. These tools are not privacy-oriented, so not all the traffic generated
by a host is tunneled, but only additional network interfaces are created to
communicate with other nodes on the same overlay network. In our implementation
we choose Nebula, the only free and open source tool suitable for the task in
the edge context.

\subsection{Container Orchestrator}

Container orchestrators are responsible for integrating, scaling, and managing
containers, while at the same time providing various functions, such as
security, networking, service discovery, and monitoring. From all available
options today, Kubernetes is the most widely adopted in large production
environments. It has the advantage of not being constrained to one specific
container runtime, but open to Docker, containerd, CRI-O, or any following the
Kubernetes Container Runtime Interface specifications. Some forks of Kubernetes
have been created, such as K3s or k0s, specifically designed for constrained
edge devices. While both K3s and k0s options are free and open source, we choose
K3s for being the most stable version at the time of implementation.

\subsection{Serverless Framework}

While containerization alone facilitates the packaging and deployment of
applications, system administrators are still in charge of configuring the
container orchestrators for scalability. The serverless model removes this need,
and system administrators are not anymore required to deal with scaling issues.
Cloud providers are already offering serverless services with solutions such as
Amazon Lambda or Google Cloud functions which provide deployment of functions
offering IDEs, SDKs, plugins, etc. Free and open source serverless platforms,
such as OpenFaaS and OpenWhisk are currently under development. In our
implementation, we use OpenFaaS which today provides a good maturity level and
fewer hardware requirements as compared to OpenWhisk.

\subsection{Serverless Functions and Workflows}

Serverless functions are pieces of code that once deployed are only executed
when they are explicitly triggered and deallocated afterwards. Typically, these
functions are single purpose, stateless and run for short periods of time. To
create more complex applications, the so-called workflows are created by
chaining functions. Chaining functions can be performed from the client side or
from the server side. From the client side, the client concatenates the output
from one function to the input of subsequent one and every function is
independently triggered. This method follows the original philosophy behind
serverless functions where the developer has full control of the workflows.
However, from the performance perspective this method creates an overhead on the
communication side which may result in higher response times. To address this,
chaining can be done on the server side. This allows functions to trigger other
functions, thus reducing the communication overhead, at the expense of client
losing control of the workflows.

\begin{figure*}[!t]
    \centering
    \includegraphics[width=0.9\textwidth]{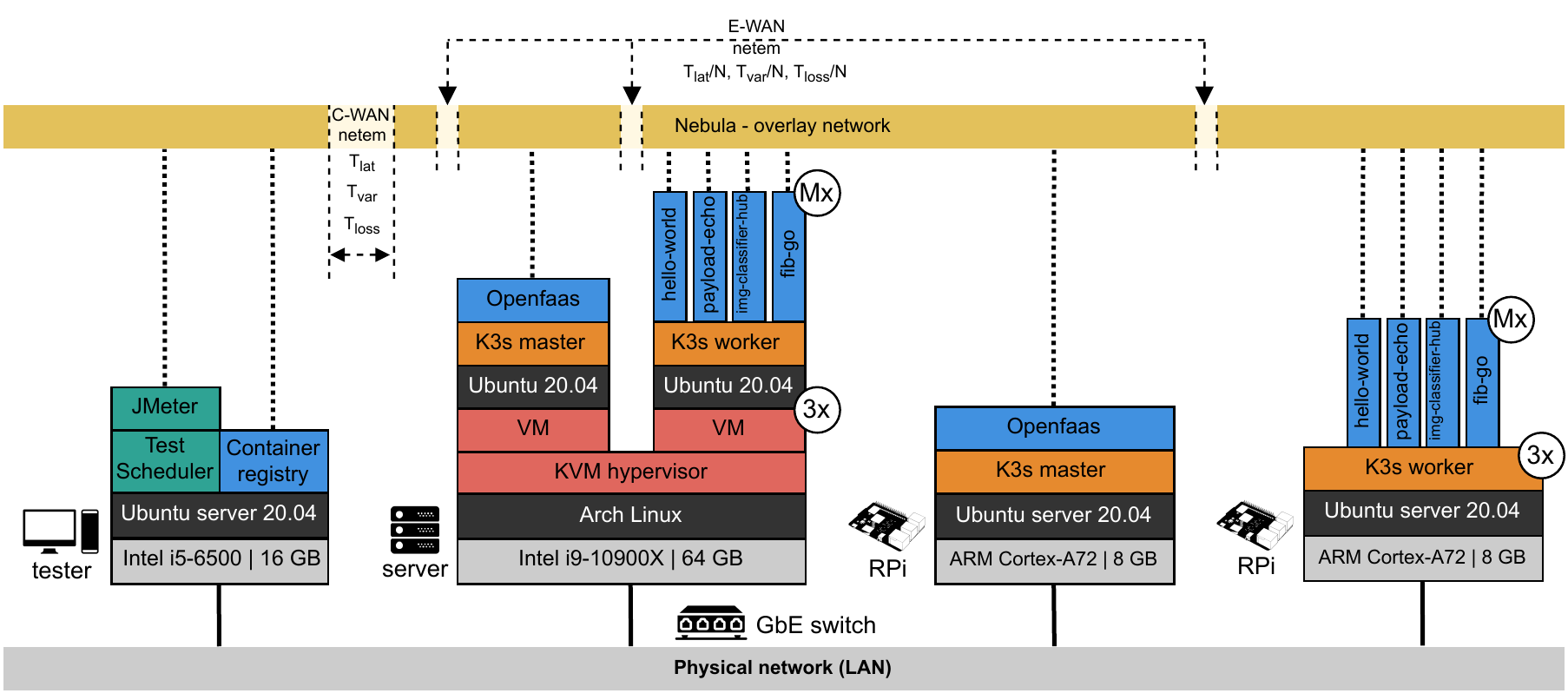}
    \caption{\testbedname~specific testbed for experimental benchmarking.}
    \label{fig:testbed}
\end{figure*}

\section{Testbed Deployment and Configurations}

\subsection{Testbed Deployment}

Fig. \ref{fig:testbed} shows the specific hardware configuration we use in our
testbed to evaluate the performance of different combinations of VMs and PMs, as
well as the software stack. More in detail, it consists of one tester machine,
from where performance tests are launched, one server, where the VMs are
deployed, and 4 Raspberry Pis (\rpmetal), working as PMs; all connected to a GbE
switch. The tester is a regular desktop computer with an Intel i5-6500 CPU and
16 GB of RAM memory running Ubuntu server 20.04. The \emph{Testbed Controller}
executes the serverless function tests using JMeter with cookies, and it
disables \emph{keep-alive} to prevent the re-use of TCP connections.
\emph{Testbed Controller} also considers the case where a local registry
contains all the container images necessary for the deployment. The performance
tests are launched to a specific cluster configuration (e.g., 2 VMs or 3 PMs).
The server with an Intel i9-10900X CPU and 64 GB of RAM memory runs Arch Linux
and deploys a KVM hypervisor where up to 4 VMs running Ubuntu 20.04 are
allocated. Three different sizes of VMs can be configured: \smallvm~(1vCPU, 2GB
of RAM), \mediumvm~(2vCPUs, 4GB of RAM) and \largevm~(4vCPUs, 8GB of RAM). All
VMs are clustered using K3s where, in all cases, one acts as a master and the
other two as workers. On K3s master, all containers related to OpenFaaS are
deployed, while the serverless functions are to be deployed only on the worker
nodes. Similarly, Raspberry Pis are also clustered with the same configuration
as the VMs, so one master node and up to three workers. We deploy version 4,
with an ARM Cortex-A72 CPU and 8 GB of RAM running Ubuntu server 20.04.

\subsection{Network Configurations}

All VMs and PMs are connected by an overlay network created using Nebula which
creates specific virtual interfaces. On these interfaces, different WAN
properties, specifically delay, variance and packet loss, are emulated by making
use of the \texttt{netem} tool, included in the Linux Kernel. It should be noted
that additional WAN network properties, such as bandwidth which we do not
consider in this paper, could also be configured as long as they are supported
by \texttt{netem}. We distinguish between centralized or distributed cluster
configuration depending on how these emulated WAN values are applied. When WAN
values are only added in between the tester machine and the rest of the cluster
(i.e., C-WAN values, see Fig. \ref{fig:testbed}), the testbed emulates a typical
cloud scenario, where the cluster is centralized at a certain distance from the
end-user. When WAN values are added in between the K3s nodes in the cluster
(i.e., E-WAN values), the testbed emulates an edge scenario where the cluster is
distributed over different locations. According to the typical network latencies
reported in \cite{fcc2021, opensignal2019}, and the typical latency values
measured for cloud and edge data centers \cite{charyyev2020}, we can define five
different network scenarios to emulate: local (\lcl), cloud (\cld), edge worst
case (\wst), edge typical case (\typ) and edge optimal case (\opt). In the local
scenario (\lcl), no delay values are added to the overlay network at all, so
this case is used as reference point. The cloud scenario (\cld) adds C-WAN
values (see Fig. \ref{fig:testbed}) which correspond to the latency values that
follow a normal distribution with a mean value of $T_{lat}=25ms$, variance
$T_{var}=5ms$ and packet loss of $T_{loss}=0.4\%$. For the three edge cases
(\wst, \typ~and \opt) E-WAN values are used and defined as $T_{lat}/N$,
$T_{var}/N$ and $T_{loss}/N$, respectively. Specifically, in the edge worst case
(\wst) $N=2$, so that the end-to-end delay values applied from the tester to any
node are equivalent to the cloud case. The edge typical case (\typ) is
configured to perform with more typical latency values, i.e., $N=3$. The optimal
edge case (\opt) is the best case scenario which, by considering real world
measured values \cite{charyyev2020}, would be equivalent to latencies below 10
ms, i.e., $N=5$. 

\begin{figure}[!t]
    \centering
    \includegraphics[width=0.9\columnwidth]{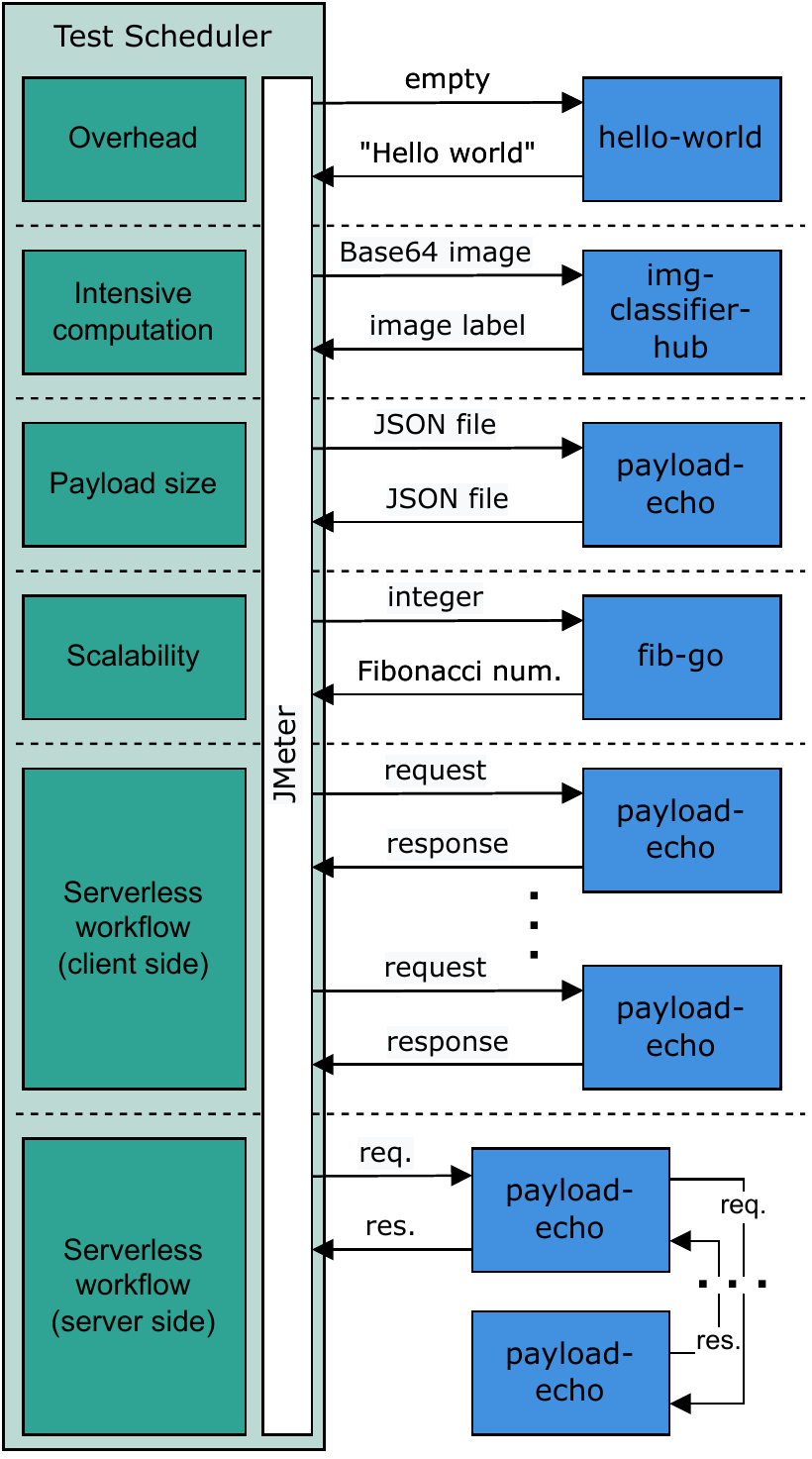}
    \caption{Test Scheduler and serverless functions}
    \label{fig:tests}
\end{figure}

\subsection{Performance Tests and Serverless Functions}

Once the testbed is deployed, including hardware and network configurations, the
\emph{Test Scheduler} launches different tests using JMeter (see Fig.
\ref{fig:tests}) with four serverless functions: \texttt{hello-world},
\texttt{img-classifier-hub}, \texttt{payload-echo} and \texttt{fib-go}. In the
first test, which we call \emph{Overhead}, we trigger \texttt{hello-world}
function to evaluate the impact on the response time caused by the testbed
infrastructure itself. In this function, single thread test performs a certain
number of requests with empty payload to the \texttt{hello-world} function. In
the second test, called \emph{Intensive computation}, we use
\texttt{img-classifier-hub}, where a single thread performs a certain number of
requests sending an encoded Base64 image as payload to the function. This
function then decodes and classifies the image, making use of the pre-trained
\emph{Inception v3 model}, a convolutional neural network,  and returns the
corresponding label for the image. For testing purposes, the image we send is
always the same one in order to avoid different decoding times. In the third
test called \emph{Payload size}, we measure the impact of the payload size on
the response times when triggering serverless functions. To that end, we use
\texttt{payload-echo} function which returns the exact same JSON file received
with no further processing. In the fourth test, called \emph{Scalability}, we
use \texttt{fib-go} function, which calculates the Fibonacci number accordingly
and returns the value, so the higher the integer value sent, the longer the
processing time. In the last two tests, \emph{Serverless workflow}, we again use
\texttt{payload-echo} function, but this time to analyze function chaining. This
test considers two operation modes, one for chaining functions on the client
side and one for chaining on the server side. For chaining on the client side,
JMeter triggers certain number of \texttt{payload-echo} functions sequentially
and the elapsed time here considers the total time from when the first function
is triggered until the response of the last function is received. For chaining
on the server side, the process of triggering functions is nested on the server
side, so functions themselves trigger other functions of the same type. The
elapsed time here considers the response time of only the initially triggered
function from JMeter side, since this one finishes after all functions have
finished.

\definecolor{tableblack}{HTML}{636363}
\definecolor{tableblackdark}{HTML}{333333}
\definecolor{tablered}{HTML}{840000}
\definecolor{tablereddark}{HTML}{4d0000}
\definecolor{tablegreen}{HTML}{496933}
\definecolor{tablegreendark}{HTML}{243419}
\definecolor{tableblue}{HTML}{3D5F9A}
\definecolor{tablebluedark}{HTML}{1d2d49}
\definecolor{tableyellow}{HTML}{999900}
\definecolor{tableyellowdark}{HTML}{4d4d00}

\begin{table*}[!t]
    \begin{center}
            \begin{tabular}{l l c c c c c}
                    \toprule
                     &  & loc & cld & ewst & etyp & eopt \\
                    \midrule
                    overhead & rpi.metal & \color{tableblackdark}14.0 - 3.0 & \color{tableyellowdark}40.0 - 6.0 & \color{tablereddark}40.0 - 4.0 & \color{tablebluedark}31.0 - 3.0 & \color{tablegreendark}24.0 - 3.0 \\
                     & vm.large & \color{tableblack}9.0 - 2.0 & \color{tableyellow}35.0 - 4.0 & \color{tablered}36.0 - 3.0 & \color{tableblue}27.0 - 2.0 & \color{tablegreen}20.0 - 2.0 \\
                     \midrule
                    intensive & rpi.metal & \color{tableblackdark}659.0 - 57.0 & \color{tableyellowdark}920.0 - 100.0 & \color{tablereddark}705.0 - 31.0 & \color{tablebluedark}686.0 - 29.0 & \color{tablegreendark}671.0 - 23.0 \\
                     & vm.small & \color{tableblack}181.0 - 9.0 & \color{tableyellow}429.0 - 80.0 & \color{tablered}241.0 - 15.0 & \color{tableblue}214.0 - 12.0 & \color{tablegreen}196.0 - 8.0 \\
                     & vm.medium & \color{tableblack}189.0 - 20.0 & \color{tableyellow}436.5 - 81.0 & \color{tablered}241.0 - 29.0 & \color{tableblue}220.0 - 23.0 & \color{tablegreen}202.0 - 21.0 \\
                     & vm.large & \color{tableblack}176.0 - 12.0 & \color{tableyellow}423.0 - 79.0 & \color{tablered}232.0 - 23.0 & \color{tableblue}211.0 - 19.0 & \color{tablegreen}195.5 - 14.0 \\
                     \midrule
                    payload 1KB & rpi.metal & \color{tableblackdark}18.0 - 2.0 & \color{tableyellowdark}44.0 - 5.0 & \color{tablereddark}44.0 - 3.0 & \color{tablebluedark}35.0 - 3.0 & \color{tablegreendark}28.0 - 2.0 \\
                     & vm.large & \color{tableblack}11.0 - 2.0 & \color{tableyellow}37.0 - 5.0 & \color{tablered}38.0 - 3.0 & \color{tableblue}29.0 - 2.0 & \color{tablegreen}22.0 - 1.0 \\
                     \midrule
                    payload 10KB & rpi.metal & \color{tableblackdark}26.0 - 3.0 & \color{tableyellowdark}62.0 - 5.0 & \color{tablereddark}64.0 - 4.0 & \color{tablebluedark}50.0 - 3.0 & \color{tablegreendark}39.0 - 3.0 \\
                     & vm.large & \color{tableblack}13.0 - 1.0 & \color{tableyellow}55.0 - 5.0 & \color{tablered}56.0 - 3.0 & \color{tableblue}41.0 - 2.0 & \color{tablegreen}29.0 - 2.0 \\
                     \midrule
                    payload 100KB & rpi.metal & \color{tableblackdark}74.0 - 7.0 & \color{tableyellowdark}359.0 - 82.0 & \color{tablereddark}208.0 - 15.0 & \color{tablebluedark}153.0 - 18.0 & \color{tablegreendark}112.0 - 11.0 \\
                     & vm.large & \color{tableblack}26.0 - 4.0 & \color{tableyellow}321.0 - 73.0 & \color{tablered}181.5 - 15.0 & \color{tableblue}121.0 - 11.0 & \color{tablegreen}76.0 - 13.0 \\
                     \midrule
                    payload 1000KB & rpi.metal & \color{tableblackdark}529.0 - 81.0 & \color{tableyellowdark}2084.0 - 503.0 & \color{tablereddark}796.0 - 122.0 & \color{tablebluedark}677.0 - 96.0 & \color{tablegreendark}612.0 - 79.0 \\
                     & vm.large & \color{tableblack}118.0 - 23.0 & \color{tableyellow}1904.0 - 462.0 & \color{tablered}592.0 - 86.0 & \color{tableblue}429.0 - 53.0 & \color{tablegreen}291.0 - 31.0 \\
                    \bottomrule
            \end{tabular}
    \end{center}
    \caption{Overhead, intensive and payload size results: median and IQR of response time in ms.}
    \label{table:results_123}
\end{table*}

\section{Measurements}

In this section, we show the measurements and the results obtained for each
benchmarking test and for each pair of cluster configurations and network
scenarios. To reduce bias due to the dynamic state of the allocated resources
managed by the operating system, the results shown consider 10 repetitions for
each test/hardware setup.

We use the previously defined tests, i.e.,  overhead, intensive computations,
payload size, scalability and serverless workflows.  Each test uses one or more
threads, where each thread performs a certain number of HTTP requests to
serverless functions specifically implemented for this purpose. Due to nature of
the tests defined, the HTTP requests are performed at constant rate so that the
performance of the system and the effect of the network can be better
determined. It should be noted, however, that JMeter also allows for other
distributions to be used by modifying accordingly the JMeter test plan files.
All serverless functions have been implemented using the production-ready
guidelines from OpenFaas and are publicly available on Github and Docker
Hub\footnote{F. Carpio, “fcarp10/openfaas-functions,” GitHub, 2022. [Online].
Available: https://github.com/fcarp10/openfaas-functions}.

\subsection{Overhead}

This measurement evaluates the impact on the response time caused by the testbed
infrastructure itself, i.e., the performance offset. This test is used as a
reference point to be able to compare the results of other tests. This single
thread test performs 100 requests at 5 req/s to the \texttt{hello-world}
function. Since this test does not require much processing, the hardware
configurations considered only compare \rpmetal with \largevm, which both
allocate the same amount of RAM. In this case, the cluster is configured with
one master and one worker node. Table. \ref{table:results_123} shows the median
and IQR values of the response times in milliseconds of the overhead test.
Starting with the local case (\lcl), we see how the RPs perform around 5 ms
slower than the VMs. When comparing \cld~cases, we can see the results are
shifted by around 25 ms with respect to the local cases, but with a higher
variance as expected due to the emulated WAN values. When comparing the edge
cases, we can see how \wst~performs similar to \cld, which is as expected since
the emulated end-to-end WAN values are equivalent in both cases. We can also
verify how \typ~and \opt~cases have lower response times as expected, but in all
cases, VMs perform slightly faster than RPs.

\subsection{Intensive computations}

A single thread test performs 100 requests at 0.5 req/s at constant rate sending
an encoded Base64 image as payload to the \texttt{img-classifier-hub} function.
Since this test requires intensive processing, we analyze here RPs and all three
VM sizes. At the same time, the cluster is only created with one master and one
worker, since this function does not perform concurrent requests. Table
\ref{table:results_123} shows the obtained results for RPs and all three VM
sizes. We can see how the performance of RPs is considerably worse than of any
VM size, even though RPs have as much RAM as the large VMs. This shows the CPU
limitations of RPs when requiring more intensive computing tasks. When comparing
different network configurations, we can see how \cld~case performs considerably
worse than any of the \emph{edge} cases. This can be explained by the fact that
sending a large payload, in this case the image to classify, over link with a
higher latency directly interferes with TCP performance. This is not the case in
the edge context where even with the comparable end-to-end latency, the master
node is closer to the tester machine. Comparing the results with respect to the
VM sizes shows a slightly improved response time with the increasing VM size,
but the improvement is negligible.

\subsection{Payload size}

This test runs 100 requests at 0.2 req/s where each request sends a JSON file as
payload to the \texttt{payload-echo} function. We consider different sizes (1,
10, 100 and 1000 KB of payload) for the JSON files. This test does not require
much processing, so only RPs are compared to VM large, using in both cases one
master and one worker. Table \ref{table:results_123} shows the results obtained.
When sending 1 KB, both RP and VM perform quite similar, with the VM being
slightly faster overall. When comparing cloud and edge cases, we can see how the
results are equivalent to the overhead test, albeit with slightly slower
response times due to the higher payload size. With 10KB, the difference between
each case becomes more evident with higher latencies overall. Moving to 100 KB
case, we see how \cld~case here, performs worse than any of the edge cases,
specially when using RPs. This behavior is even more obvious with 1000 KB of
payload, even when using the large VM. As in the previous test, while both
\cld~and \wst~cases have the same emulated WAN values, the cloud performance is
penalized by the higher latency to the master node, which directly interferes
with TCP performance.

\begin{figure*}[!t]
    \centering
    \subfloat[Fib. num. 1]{\includegraphics[width=0.45\textwidth]{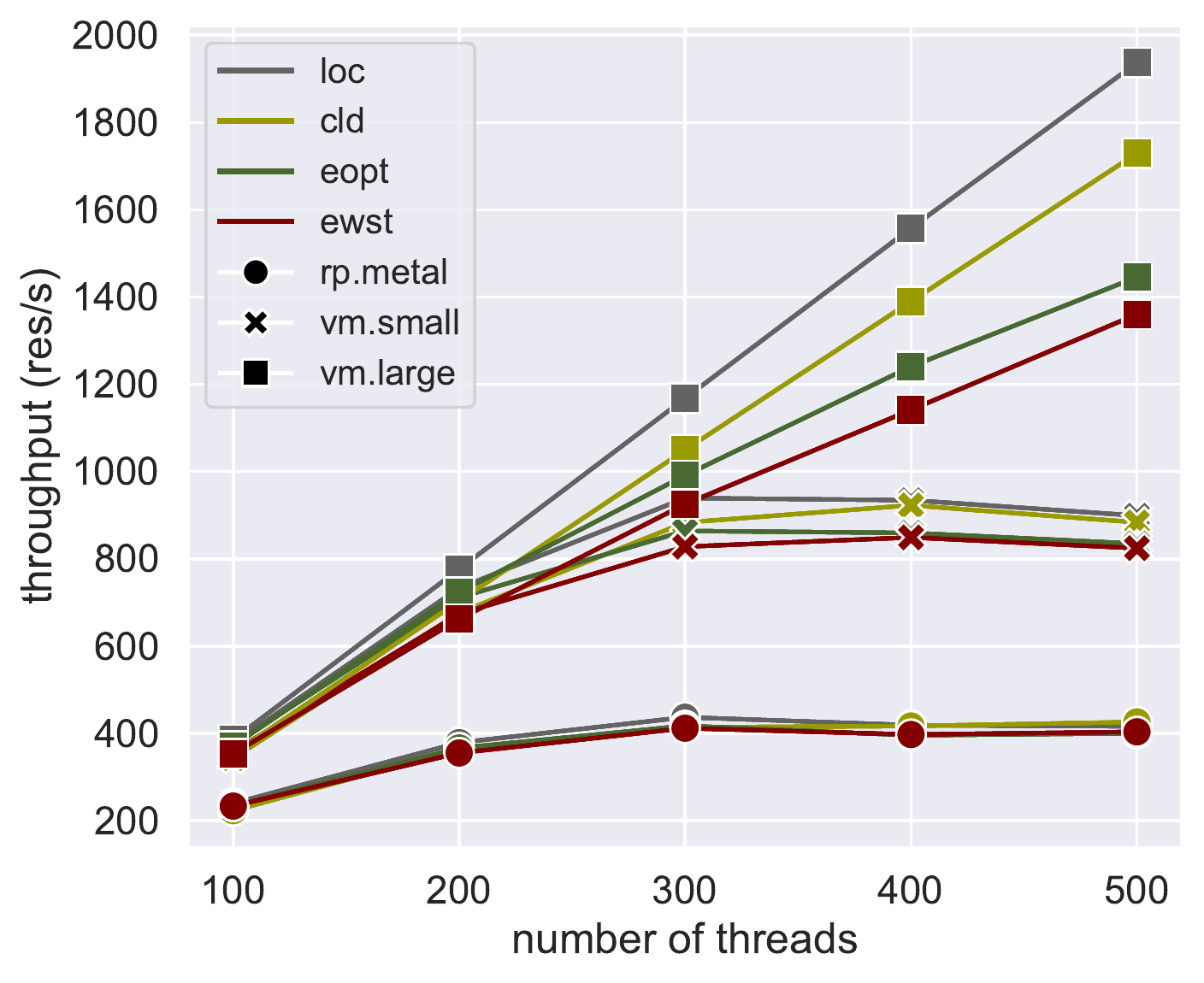}
        \label{fig:4_scalability_1_throughput}}
    \hfil
    \subfloat[Fib. num. 30]{\includegraphics[width=0.45\textwidth]{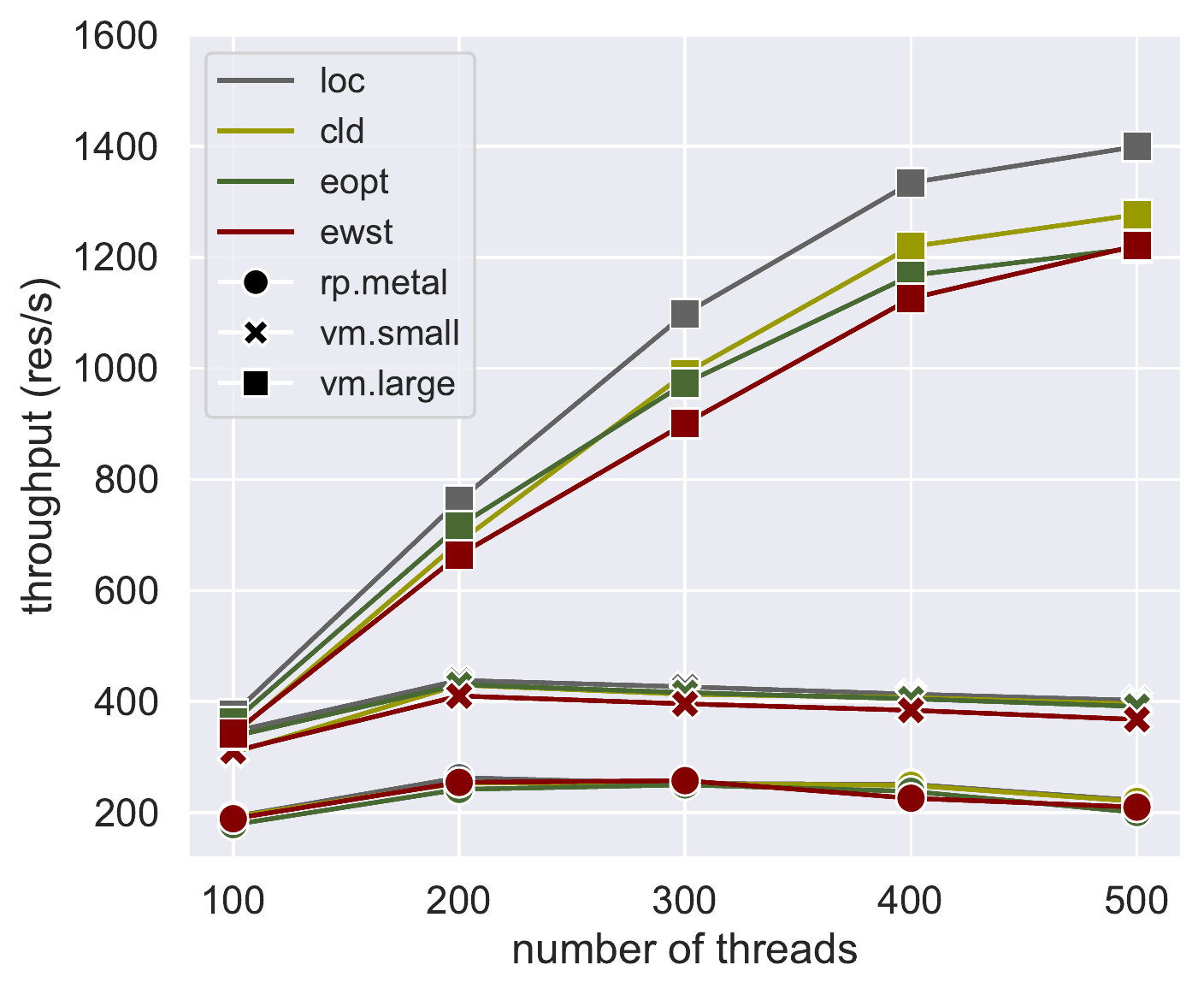}
        \label{fig:4_scalability_30_throughput}}
    \caption{Scalability - throughput in number of successful req/s.}
    \label{fig:4_scalability}
\end{figure*}

\begin{figure*}[!t]
    \centering
    \subfloat[Workflow of lenght 5]{\includegraphics[width=0.48\textwidth]{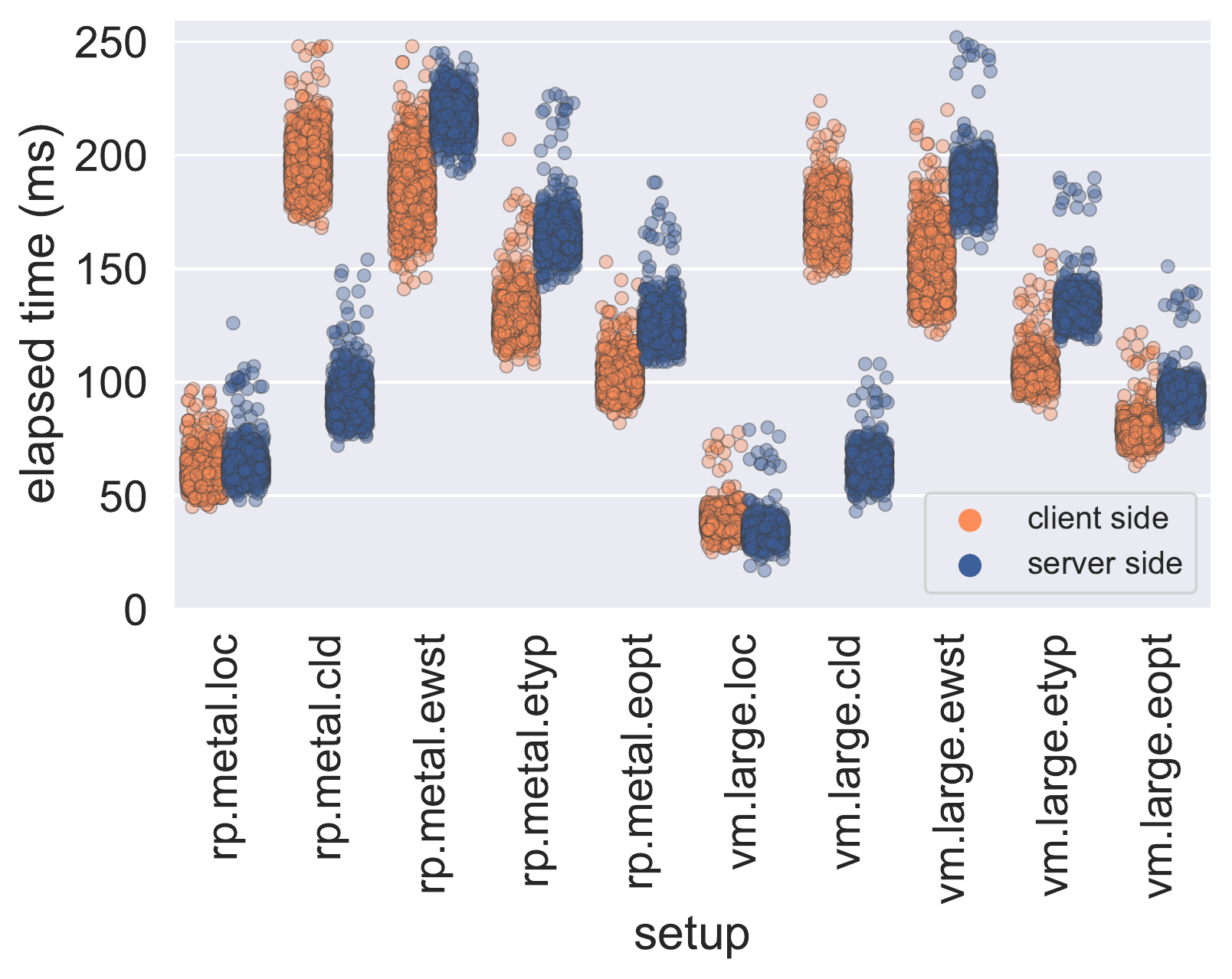}%
        \label{fig:5_chains_5}}
    \hfil
    \subfloat[Workflow of lenght 20]{\includegraphics[width=0.48\textwidth]{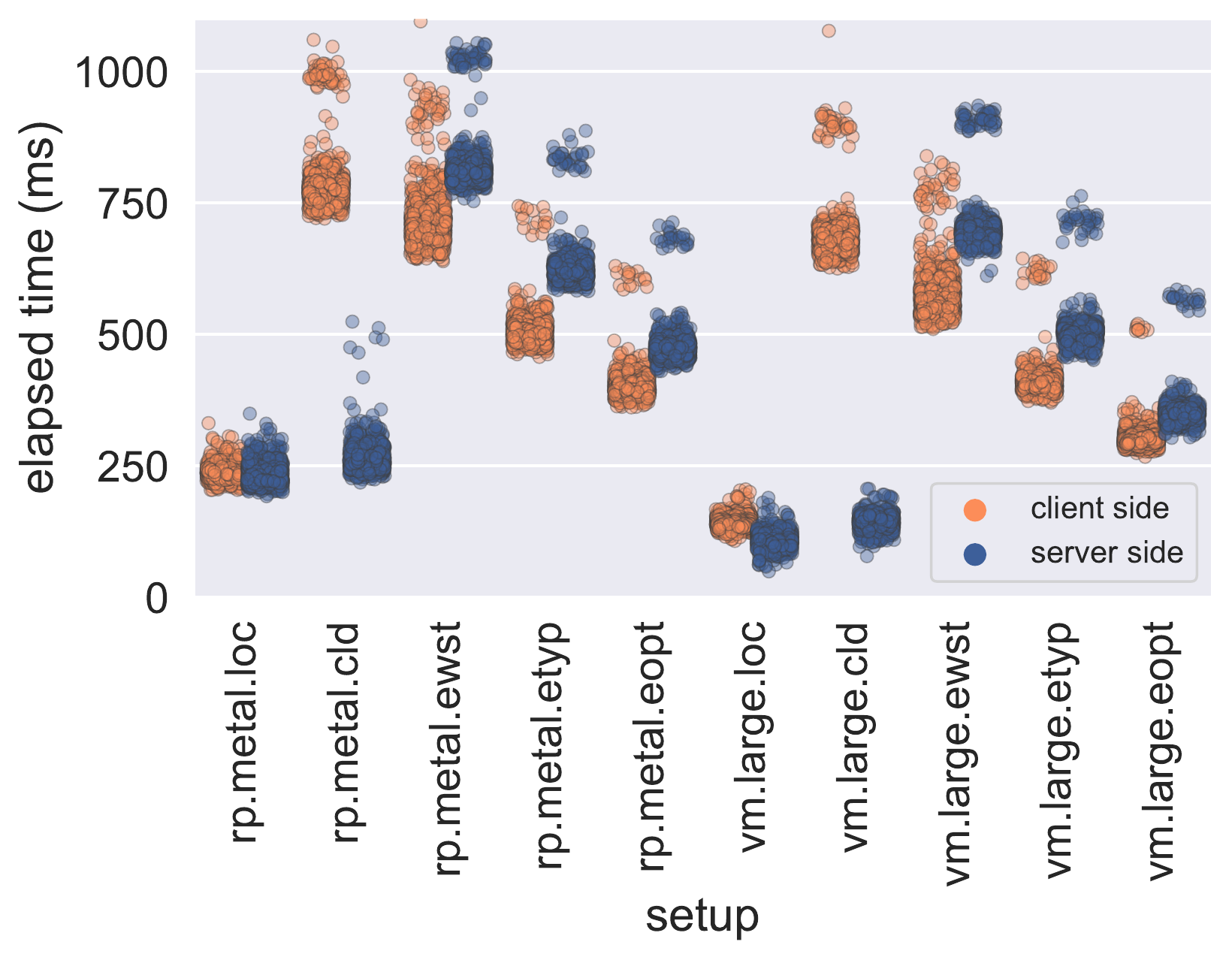}%
        \label{fig:5_chains_20}}
    \caption{Serverless workflows - elapsed time  in ms.}
    \label{fig:chains_results}
\end{figure*}

\subsection{Scalability}

We evaluate scalability by opening multiple threads where each thread performs 1
request every 250 ms.  In every request, an integer is specified to the
\texttt{fib-go} function. The number of threads span from 100 to 500, meaning
the system receives between 400 and 2000 req/s. The duration of the test is set
to 5 minutes. Since this test is designed to stress the system, the cluster is
set up with one master and three workers, using different VMs sizes and RPs. We
show results comparing when (i) all threads are requesting for the Fibonacci
number 1, so no extra processing required in the function, to (ii) when the
requested Fibonacci number is 30, so that a larger processing load is required.

Fig. \ref{fig:4_scalability_1_throughput} shows the throughput for different
number of threads comparing when asking for the Fibonacci number 1. Here, we can
see already how RPs perform worse than any VM size with the performance peak at
300 threads delivering around 400 successful responses per second. Also, the
\smallvm~achieves the maximum performance at 300 threads with around 900
responses per second. The performance of \largevm~ is not limited in this case
by the number of threads, but increases linearly. There are also no larger
differences in performance between \lcl~and \cld~cases, specially when the
system delivers maximum throughput. We also notice here that there are no major
differences between \wst~and \opt~cases, but in both cases the throughput is
much lower compared to the \cld~case. This shows how the cloud case performs
much better overall than any edge case, achieving up to 1750 res/s in the best
case scenario as compared to around 1400 res/s in the edge case. This is likely
because of the intra-communication in the cluster that does not experience
performance penalties as large as in the edge. When increasing the Fibonacci
number to 30, see Fig. \ref{fig:4_scalability_30_throughput}, the overall
throughput decreases, as expected. Here, we can better observe the impact of the
network specially for \largevm~where \lcl~case performs much better than \cld~,
but not much difference when compared to \opt~and \wst~since the processing
time is the limiting factor. 

\subsection{Serverless Workflow}

We now provide results when chaining serverless functions using the
\texttt{payload-echo} function. The serverless workflow test performs 100
iterations at 10 req/s, considering one master and 3 worker nodes configured
with either \rpmetal~or \largevm. Fig. \ref{fig:5_chains_5} shows the strip
plots of response time when chaining the \texttt{payload-echo} function 5 times
from the client and the server sides, respectively. Here, we see how \cld~case
is clearly affected when chaining on the client side compared to the server
side. This is expected, due to the emulated WAN values which are only present in
between the tester and the cluster, in \cld~case, making the chaining on the
server side more advantageous. This is not the case, however, with any of the
edge cases since emulated WAN values are equally applied to all nodes, so
chaining on the server side actually increases the response time, since
functions may trigger other functions that are deployed on different nodes, so
the intra-latencies are penalizing the performance. The results are also
coherent when increasing the number of functions to 20, see Fig.
\ref{fig:5_chains_20}, where all response times increase by around a factor 4,
as expected. An important aspect to notice here is that while the performance of
VMs is slightly better than RPs, the difference is not as remarkable, since
there are no CPU intensive tasks involved in this test.

\subsection{Discussion and Remarks}

When comparing the performance of RPs vs VMs, we can see how RPs are as capable
as VMs in tasks that do not require compute intensive tasks, but there is a
clear limitation on the maximum throughput that can be achieved when scaling
functions. The size of VMs also affects scaling, where the number of allocated
resources limits the throughput of the system. On the one side, considering that
serverless systems are, in general, not intended to be used with functions that
require long computation times, in most of the cases we can say that RPs are
appropriate. On the other side, since scalability is the key feature in
serverless systems where a high number of concurrent requests is expected, the
system needs to be properly dimensioned. With proper dimensioning of the system,
RPs can be as performant as VMs in most of the use cases that the serverless
paradigm is targeting.

Regarding the impact of network, we find that having a distributed cluster is,
at first, suboptimal due to the intra-communication caused by the cluster nodes.
In some cases, however, this effect is negligible compared to the benefits of
having shorter distances between nodes. This is obvious when the function
requires transferring large amounts of data, in which case the edge network
system outperforms the cloud. However, considering that serverless functions are
not intended for usage with large payloads, this advantage can be questioned,
and it will depend on application design. The effect of the network is also
irrelevant when the processing time is too high due to either limited resources
or lack of proper scalability. We observed that distributed computing in the
network is not always a better choice; instead, a proper, even a joint design of
system and application maybe needed.

The main value of this testbed for users is in its reproducibility of the
 infrastructure and the results obtained, since it does not rely on either third
 party cloud services, the state of the real-world network or the Internet
 bandwidth. In addition, it is easy to automatically deploy, since new
 performance tests can be added in form of JMeter files together with new
 functions; once implemented, the rest of the system does not need to be
 reconfigured. At the same time, the testbed has its limitations. For instance,
 the impact of the programming language chosen for the functions is unknown,
 which maybe an important factor. We also have not tested the impact of
 bandwidth when analyzing the performance on serverless functions, which might
 be another important factor specially when sending higher payload sizes. Also,
 we did not optimize scalability, which means that we chose the threshold values
 for scaling functions based on experiments that provide the best results. This
 implies that the results in absolute values may actually get better by better
 adjusting the scaling parameters.

\section{Conclusions}

This paper aimed at contributing to community-driven experimental studies in
edge computer network systems, and shared lessons learnt in engineering a free
and open source edge computing network testbed to benchmark serverless
functions. The measurements indicated that edge computing network systems are
valid candidates for serverless functions. Future work includes adding further
performance tests to cover a broader range of serverless function scenarios as
well as on working on optimizing the scaling mechanisms. 

 \section*{Acknowledgments}
The authors acknowledge the financial support by the Federal Ministry of
Education and Research of Germany in the program of "Souverän. Digital.
Vernetzt." Joint project 6G-RIC, project ID: 16KISK031. This project has also
received funding from the European Union's Horizon 2020 research and innovation
programme under grant agreement No 952644.
\bibliographystyle{IEEEtran}
\bibliography{ref}


\newpage





\vfill

\end{document}